\documentclass[ reprint, prl, superscriptaddress, showpacs, preprintnumbers]{revtex4-1}
\usepackage{amsmath, amssymb}
\usepackage{textcomp}
\usepackage[dvips]{color,graphicx}

\usepackage{graphicx}% Include figure files
\usepackage{bm}% bold math
\usepackage{hyperref}% add hypertext capabilities
\begin{document}

\preprint{APS/123-QED}

\title{Pressure Tuned Insulator to Metal Transition in Eu$_2$Ir$_2$O$_7$}% Force line breaks with \\

\author{F. F. Tafti}
\affiliation{The Department of Physics, University of Toronto, 60 St. George Street, Toronto, Ontario, M5S 1A7, Canada}
\author{J. J. Ishikawa}
\affiliation{Institute for Solid State Physics, University of Tokyo, Kashiwa, Chiba 277-8581, Japan}
\author{A. McCollam}
\affiliation{High Field Magnet Laboratory, Institute for Molecules and Materials, Radboud University Nijmegen, 6525 ED Nijmegen, The Netherlands}
\author{S. Nakatsuji}
\affiliation{Institute for Solid State Physics, University of Tokyo, Kashiwa, Chiba 277-8581, Japan} 
\author{S. R. Julian}
\affiliation{The Department of Physics, University of Toronto, 60 St. George Street, Toronto, Ontario, M5S 1A7, Canada}
\affiliation{Canadian Institute for Advanced Research,
    Quantum Materials Program,
    180 Dundas St.\ W., Suite 1400, Toronto, ON, Canada M5G 1Z8}

\date{\today}% It is always \today, today,

\begin{abstract}

We have studied the effect of pressure on the pyrochlore iridate Eu$_2$Ir$_2$O$_7$, which at 
ambient pressure has a thermally driven insulator to metal transition at $T_{MI}\sim120$\,K. 
As a function of pressure the insulating gap closes, apparently continuously, near $P \sim 6$\,GPa. 
However, rather than $T_{MI}$ going to zero as expected, the insulating ground state crosses 
over to a metallic state with a negative temperature coefficient of resistivity, 
%%%%%%%%%%%%%%%%%% original %%%%%%%%%%%%
%% calling into question the true nature of both ground states. 
%%%%%%%%%%%%%%%%%%% replacement %%%%%%%%%%
suggesting that these ground states have a novel character.
%%%%%%%%%%%%%%%%%%%%%%%%%%%%%%%%%%%
The high temperature state also crosses over near 6 GPa, from an incoherent to a 
  conventional metal, implying that there is a connection between the high and the low 
  temperature states.

\end{abstract}

\maketitle

\section{Introduction}

%%%%%%%%%%%%%%%%%  my replacement %%%%%%%%%%%%%%%%%%%%%%
Putting itinerant electrons with strong spin-orbit coupling on 
  geometrically frustrated lattices offers new possibilities for 
  strongly correlated electron states. 
Recent attention has focused on the pyrochlore iridates, 
  R$_2$Ir$_2$O$_7$ where R is a rare earth, for which  
  several theoretical studies propose the existence of 
  topologically non-trivial ground-states 
  \cite{pesin10,wan11,kim11,hosur11}.
Testing such predictions requires advanced experiments, 
  for example measurements that can reveal 
  unconventional quantum phase transitions \cite{tewari11}.  
Here, we report an unusual insulator-to-metal quantum phase 
  transition in 
  the pressure-temperature phase diagram of the pyrochlore iridate 
  Eu$_2$Ir$_2$O$_7$. 

Amongst pyrochlore oxides, iridates have their M-sites occupied by Ir$^{4+}$ 
  ions whose extended $5d$ orbitals are prone to strong electron-lattice and 
  spin-orbit couplings \cite{korneta10}. 
Compared with the less extended orbitals of $3d$ transition metal oxides, 
  iridates are naively expected to be less strongly correlated. 
Nevertheless, the strong spin-orbit and electron-lattice couplings can lift 
  the orbital degeneracy of iridium $5d$ electrons and narrow their bandwidths. 
Hence iridates can be delicately poised near a bandwidth controlled metal 
  insulator transition (MIT). 

Successive replacement of the R site of R$_2$Ir$_2$O$_7$ with larger rare earth 
  atoms causes a change from insulating to metallic behaviour 
  \cite{yanagashima01, matsuhira07}.
Progressing from right to left in the Lanthanide series, 
  (Lu, Yb, ..., Gd)$_2$Ir$_2$O$_7$ are all insulators;
  (Eu, Sm, and Nd)$_2$Ir$_2$O$_7$ are on the boundary, looking 
  metallic at high temperatures but insulating at low temperatures, 
  and Pr$_2$Ir$_2$O$_7$ is metallic down to the 
  lowest temperatures. 
Increasing the R size in the R$_2$Ir$_2$O$_7$ series makes the Ir-O-Ir bond 
  angle wider and the Ir-O bond length shorter.
As a result, the iridium $t_{2\textrm{g}}$ bandwidth increases and eventually 
  passes the metallization threshold beyond a certain R ionic radius \cite{koo98}. 

As the R size increases from Eu to Nd,  
  (Eu, Sm, and Nd)$_2$Ir$_2$O$_7$ show several 
  signatures of a weakening low-temperature insulating phase and a 
  strengthening high-temperature metallic phase \cite{matsuhira07}: 
(1) the metal-insulator transition temperature $T_{MI}$ is significantly 
  smaller in the Nd compound (36\,K) compared to the 
  Eu and Sm compounds (120 and 117\,K respectively);
(2) the small and temperature dependent 
  gap ($\Delta<10$\,meV) of the insulating phase 
  is smallest in the Nd compound  \cite{matsuhira07};
(3) the Nd compound has the smallest inverse $RRR$ ratio defined as 
  $R_{4\text{K}}/R_{300\text{K}}$;  
  and 
(4) in the high temperature metallic phase only the resistivity of 
  the Nd compound is `metallic', having a positive slope for $\rho(T)$; 
  in Eu and Sm $\rho(T)$ is `non-metallic', with a weakly negative slope.  
(By `non-metallic' we mean a resistivity that rises with decreasing temperature,  
  but with power-law temperature dependence as opposed to the exponential 
  temperature dependence of a gapped insulator.) 

These three systems are of particular interest at present, because 
  it is believed that states near the metal-insulator boundary in the 
  pyrochlore iridates may 
  have a toplogical nature 
  \cite{pesin10,wan11,kim11,hosur11}.  
In particular, Wan et al.\ \cite{wan11} and Witczack-Krempa et al.\ \cite{kim11} 
  predict that a toplogical semi-metallic state separates the insulating  
  and the metallic ground states. 
This `Weyl semi-metal' is a three-dimensional analogue of graphene, with the 
  chemical potential pinned to Dirac points of chiral states. 
In order to form, the Weyl semi-metallic state  requires magnetic order, and  
  due to the vanishing density of states at the chemical potential, the electrical 
  conductivity of the bulk is predicted to vanish when $T \rightarrow 0$ K 
  \cite{wan11,balents11,hosur11}.

Electron interactions and correlations of pyrochlore iridates can be tuned by 
  either chemical or physical pressure. 
For investigating the nature of the low temperature state, physical pressure has the 
  obvious advantage that it can be tuned continuously. 
(Continuous substitution studies in the pyrochlore iridates are not 
  useful, since the ground state is sensitive to disorder \cite{matsuhira07}.) 
Physical pressure is somewhat different from chemical pressure, however. 
Increasing the R atomic size increases the Ir-O-Ir bond angle and the 
  lattice parameter in parallel \cite{yanagashima01}, 
  while hydrostatic pressure increases the former but decreases 
  the latter. 
Thus we expect new insights from the application of physical 
  pressure to explore the boundary between metallic and insulating ground states 
  in pyrochlore iridates. 

%%%%%%%%%%%%%%%%%%%%%%%23 Dec 2011
\section{Experiment}
%%%%%%%%%%%%%%%%%%%%%%%

Eu$_2$Ir$_2$O$_7$ single crystals were grown at ISSP using the 
  KF-flux method \cite{millican07}. 
We pressurized samples measuring approximately $150\times100\times30\ \mu$m$^3$ 
  in a moissanite anvil cell \cite{xu00} and measured resistivity 
  as a function of temperature at several pressures in the 
  range P = 2 to 12\,GPa, using a four terminal ac method. 
The pressure medium was 7373  Daphne oil, and 
  pressure was monitored by ruby fluorescence spectroscopy at room temperature. 
A 1\,K dipping probe was used in the temperature range T = 300 to 2\,K. 
Resistivity below 2\,K and magnetoresistance at 10.01\,GPa were measured in a 
  dilution refrigerator.  
 
\begin{figure}
\centering
\includegraphics[width=3.1in]{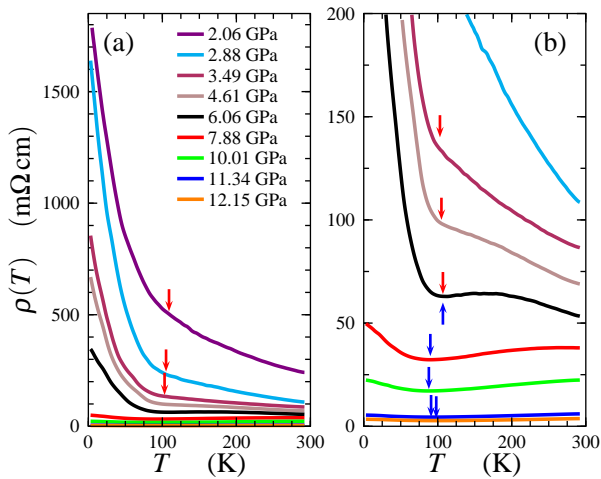}
\caption{\small 
Resistivity as a function of temperature from $P$ = 2.06 to 12.15\,GPa. 
The left-hand panel contains all of our results, showing that the resistivity 
  is strongly suppressed by increasing pressure. 
The right-hand panel focuses on the intermediate-pressure data. 
Red arrows indicate the metal-insulator transition (see Fig.\ \ref{TMI} and 
the text); blue arrows indicate the minimum in $\rho(T)$. 
}
\label{rho_vs_T}  
\end{figure}

\section{Results}

Our resistivity data, from 300 to 2\,K at nine different pressures from 
  2.06 to 12.15\,GPa, are presented in Fig.\,\ref{rho_vs_T}. 
The quantitative effects of increasing the pressure from 2 to 12 GPa are dramatic: 
  the room temperature resistivity falls by a factor of 60, 
  while 
  the 2 K resistivity falls by a factor of 500. 
Qualitatively, the slope of the resistivity, $\partial\rho(T)/\partial T$, 
  for $T>100$ K changes 
  from negative (non-metallic) at low pressure to positive (metallic) at 10 and 12 GPa. 
At lower temperature, in contrast, the slope of the resistivity 
  is negative at 
  all pressures, but it is nearly 1000 times larger at 2.06 GPa than 
  at 12.15 GPa.
We elaborate on the low temperature resistivity in the Discussion, 
  and show that the low pressure curves have a temperature dependent gap that 
  closes between 6 and 8 GPa.

\begin{figure}[t]
\centering
\includegraphics[width=3.3in]{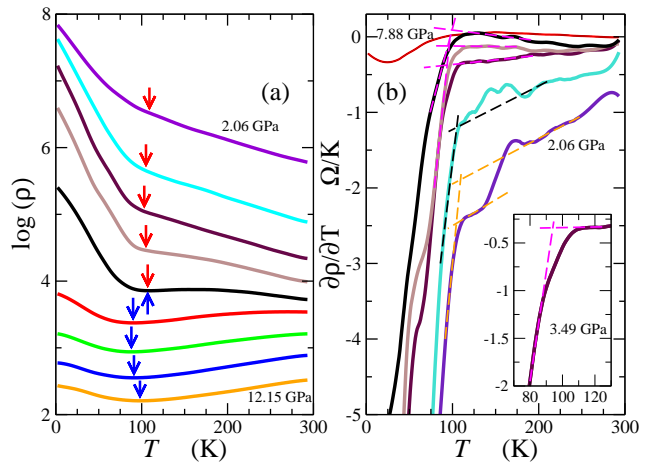} % first_deriv_2.ps}
\caption{\small Locating the metal-insulator transition, $T_{MI}$. 
The color coding is the same as in Fig.\ \ref{rho_vs_T}.
The transition does not have a strong signature in $\rho(T)$, however 
  panel (a) shows that 
  there is a marked change in slope of $\log(\rho)$ vs.\ $T$ 
  near 100 K at all pressures, while panel (b) shows that the 
  onset of this behaviour can be found, for all pressures below 
  7.88 GPa, in a sharp downturn in 
  $\partial \rho/\partial T$. 
As a guide to the eye we have shown plausible extrapolations of the 
  high-temperature slope, and the slope immediately below 100 K. 
At each pressure we have placed $T_{MI}$ at the mid-point between 
  the onset of the downturn and the temperature at which the these 
  extrapolations meet; while the error bars (see Fig.\ \ref{phasediag}) 
  extend to these two temperatures. 
The inset of (b) zooms in on $\partial\rho/\partial T$ near 100 K for 
  3.49 GPa. 
In (a) some curves are offset vertically for clarity. 
 }
\label{TMI}
\end{figure}

The metal-insulator transition, which occurs at 120 K at ambient pressure, 
  does not show up clearly in the raw resistivity. 
This is also the case at ambient pressure \cite{matsuhira07}.  
However, Fig.\ \ref{TMI} shows that, for low pressures, the slope of 
  $\log(\rho(T))$ vs $T$ changes near 100 K.  
This change appears to be quite abrupt 
  in the 3.49 GPa and the 4.61 GPa curves, 
  both in Fig.\ \ref{TMI}(a) and in the raw data 
  (Fig.\ \ref{rho_vs_T}(b)). 
This change in slope is more clearly seen in plots of 
  $\partial\rho/\partial T$ vs.\ $T$  
  (Fig.\ \ref{TMI}(b)): for all pressures below 7.88 GPa, 
   $\partial\rho/\partial T$ begins to decrease rapidly, with a 
   sharp, well-defined onset, near 100 K.  
We have used this to identify $T_{MI}$: 
  down to $T_{MI}$ the slope of $\partial\rho/\partial T$ is roughly constant, 
  then at $T_{MI}$ it begins to fall rapidly.  
The inset of Fig.\ \ref{TMI}(b) shows a clear example, 
  at 3.49 GPa. 
The red arrows in the first two figures correspond to $T_{MI}$, 
  assigned to the 
  mid-point between where the slope first starts to turn 
  downwards, and the point where the extrapolated high and 
  low temperature slopes meet.  
Error bars on our phase diagram, discussed below, extend to these two temperatures. 
The lowest pressure $\partial \rho/\partial T$ curves are rather noisy, 
  perhaps because the contact resistances, which improved as the pressure 
  increased, were rather large, but even in these cases a sharp change can be identified 
  quite accurately. 

\begin{figure}[t]
\centering
\includegraphics[width=3.2in]{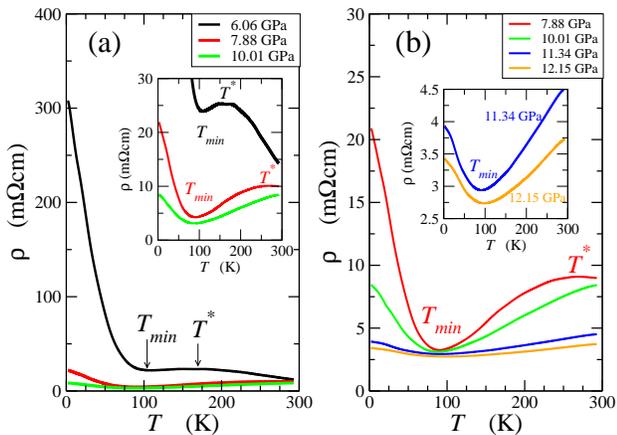}
\caption{\small Location of $T^*$ and $T_{min}$. 
The main panel in (a) shows $\rho(T)$ from 6.06 to 10.01\,GPa, 
  while the inset shows an expanded view around $T^*$ and $T_{min}$. 
  The curves are offset vertically to give a better view of the data. 
The main panel in (b) shows $\rho(T)$ from 7.88 to 12.15\,GPa. 
  The inset is an expanded view of the two highest pressure curves with 
  $RRR<1$. 
% Pressure suppresses the resistivity ratio $R_{\text{4K}}/R_{\text{300K}}$. 
}
\label{highPrhovsT}
\end{figure}

At 6.06 GPa and above, there is a minimum in $\rho(T)$, and 
  like $T_{MI}$ it is close to 100 K. 
The way this minimum develops is shown in Fig.\ \ref{highPrhovsT}. 
In the lower pressure curves the resistivity has a negative 
  slope at all temperatures, but 
  in the 6.06 and 7.88 GPa curves 
  an intemediate region of $\partial \rho/\partial T > 0$ 
  develops below a local maximum at $T^*$ and a local minimum at $T_{min}$. 
  At higher pressure the maximum has apparently moved above room temperature, 
  so $\rho(T)$ has a positive, metallic, slope from $T_{min}$ to 293 K. 
The $T^*$ cross-over occurs at 180 and 270\,K on the $P = 6.06$ and 7.88\,GPa 
 curves respectively. 

\begin{figure}
\centering
\includegraphics[width=2.75in]{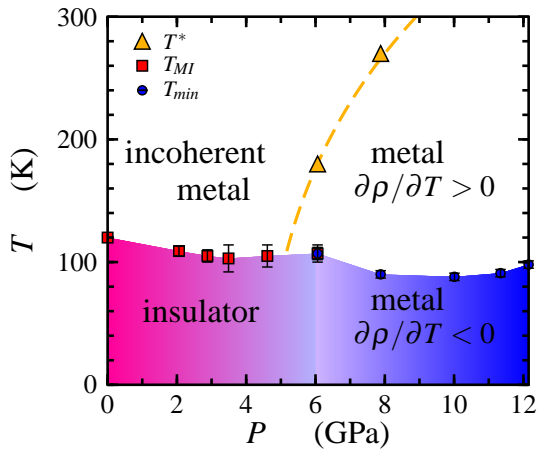}
\caption{\small The phase diagram for Eu$_2$Ir$_2$O$_7$ constructed from our 
   resistivity data. 
At low pressures, $P<6$\,GPa, the finite temperature MIT is indicated 
   by red squares; the value at 0 GPa is taken from reference \cite{matsuhira07}.  
At high pressures, $P>6$\,GPa, the transition between conventional and 
   negative-$\partial\rho/\partial T$ metallic states is indicated by blue circles. 
The $T^*$ cross-over is shown by orange triangles. 
All the lines are guides to the eye. 
  The quantum critical point (QCP) separating the insulating and the 
  negative $\partial\rho/\partial T$ metal lies on the $P$ axis near 
   $P_c=6.06\pm0.60$\,GPa.  Both $T_{MI}$ and $T_{min}$ depend weakly on $T$.}
\label{phasediag}
\end{figure}

We have used the qualitatively different resistivity behaviours  
  to construct a phase diagram, shown in Fig.\,\ref{phasediag}.  
The phase diagram can be viewed as four quadrants, corresponding to four distinct 
  regimes of electronic transport. 
In the top left quadrant ($P \lesssim 6$\,GPa, $T\gtrsim100$\,K), the system is an 
  ``incoherent metal" characterized by a high resitivity and a power-law temperature 
   dependence with a 
    non-metallic slope, $\partial \rho(T)/\partial T < 0$. 
The MIT at $T_{MI}$ separates the incoherent metallic phase from the 
  ``insulating" phase in the bottom left quadrant ($P \lesssim 6$\,GPa, $T\lesssim100$ K), 
  characterized by an exponentially activated resistivity, as discussed below. 

In the top-right quadrant ($P\gtrsim 6$\,GPa, $T\gtrsim100$\,K), the system is a 
  ``conventional metal" characterized by a metallic slope, $\partial\rho(T)/\partial T > 0$.  
The cross-over at $T_{min}$ separates the conventional metal and the 
  ``negative $\partial \rho(T)/\partial T$ metal"  
  ($P \gtrsim 6$\,GPa, $T\lesssim100$\,K), whose resistivity 
  increases with decreasing temperature in a non-Fermi liquid (NFL) power-law 
  fashion (Fig.\,\ref{highPrhovsT}(b)). 

The two high-temperature regimes, the conventional metal and incoherent metal,  
  are separated by the 
  $T^*$ cross-over, at which the slope of $\rho(T)$ changes from positive to negative.   

We distinguish between the incoherent metal (in the top-left quadrant) and the 
  negative $\partial\rho/\partial T$ metal (in the bottom right-quadrant).  
Although they both have power-law dependence on $T$ with negative $\partial\rho/\partial T$, 
  their absolute resitivities differ by a factor of up to 500,
  and the latter
  probably has Landau-type quasiparticles  
  as established by the metallic resistivity at high temperature, 
  while in the incoherent metallic phase it is likely that 
  quasiparticles have not formed. 

This phase diagram is in broad agreement with the effects of chemical 
  pressure \cite{matsuhira07}, but with the clear difference that 
  $T_{MI}$ changes rapidly with chemical pressure, 
  falling from 201 K to 36 K in going from Eu$_2$Ir$_2$O$_7$ to 
  Nd$_2$Ir$_2$O$_7$, while in our measurements it is nearly 
  pressure independent.  
Moreover, $T^*$ is not seen in the chemical pressure measurements. 

Our phase diagram suggests a connection between the high and low 
  temperature phases of Eu$_2$Ir$_2$O$_7$:
  the incoherent metal becomes insulating below $T_{MI}$;
  the conventional metal crosses over to the negative $\partial\rho/\partial T$ 
  metal below $T_{min}$; in other words 
  the transition between the insulating and the negative $\partial \rho/\partial T$ 
  metallic ground states 
  at $P_c=6.06 \pm 0.60$\,GPa coincides with the 
  incoherent-coherent cross-over at $T^*>100$\,K.

\section{Discussion}

To discuss our results, we start with the high temperature part of the phase diagram. 
The incoherent metallic phase is characterized by a 
  high resistivity and a 
  negative $\partial \rho(T)/\partial T$ that are both gradually suppressed by increasing 
  pressure. 
Metallic phases in the vicinity of localization transitions are usually subject 
  to strong fluctuations in the spin, charge and orbital degrees of freedom, 
  resulting in unconventional transport properties.
The negative $\rho(T)$ slope in the incoherent metallic regime of 
  Eu$_2$Ir$_2$O$_7$ is probably an example of such physics. 

The boundary between the two high temperature regimes is marked by the 
  dashed $T^*$ line in Fig.\,\ref{phasediag}. 
The broad peaks at $T^*$ (Fig.\,\ref{highPrhovsT}) mark a coherent-incoherent 
  cross-over of the quasiparticle dynamics that is also typical of 
  correlated oxides in proximity 
  to a Mott transition \cite{kezmsarki04,georges96}. 
The $T^*$ cross-over has not been observed in the previous chemical pressure 
  measurements in pyrochlore iridates by replacing the R site with larger atoms: 
  it presumably takes place somewhere between Sm$_2$Ir$_2$O$_7$ and Nd$_2$Ir$_2$O$_7$.  
Moreover, it probably cannot be realized by alloying on the R site, i.e.\ partially 
  replacing Eu or Sm with Nd, because of the extreme sensitivity of these systems 
  to disorder \cite{matsuhira07}. 
Physical pressure is therefore the only means by which $T^*$ can be observed. 

A similar pressure-induced coherent-incoherent cross-over has been observed 
  in Gd$_2$Mo$_2$O$_7$, which is located at the boundary between a FM metal and a 
  spin glass Mott insulator in the R$_2$Mo$_2$O$_7$ series \cite{hanasaki06}. 
Gd$_2$Mo$_2$O$_7$ goes through a continuous bandwidth-controlled insulator to 
  metal quantum phase transition at $P_c = 2.4$\,GPa 
  \cite{moritomo01}. 
Simultaneously a cross-over appears at $T^*\approx 150$ K in the resistivity 
  data. 
$T^*$ shows a strong pressure dependence in both Eu$_2$Ir$_2$O$_7$ 
  (Fig.\,\ref{phasediag}) and Gd$_2$Mo$_2$O$_7$ \cite{hanasaki06}: 
  $T^*$ shifts at a rate $\Delta T^*/\Delta P \sim 50$ K/GPa in Eu$_2$Ir$_2$O$_7$ 
  and $\sim 35$ K/GPa in Gd$_2$Mo$_2$O$_7$.

Turning now to the low temperature part of the phase diagram, 
  previous studies \cite{matsuhira07} have treated 
  the low temperature regimes of 
  Eu$_2$Ir$_2$O$_7$, Sm$_2$Ir$_2$O$_7$ and Nd$_2$Ir$_2$O$_7$ as a 
  Mott insulating phase with a so-called temperature dependent gap 
  (a gap that gets smaller with decreasing $T$, with a non-divegent 
  resistivity in the $T\to 0$ K limit). 
In all three systems there is a magnetic anomaly at the metal-insulator transition 
  \cite{matsuhira07,zhao11,taira01}.
Unlike the Sm and Nd systems, however, 
  Eu$_2$Ir$_2$O$_7$ does not show a sharp metal-insulator transition 
  in the resistivity, rather there is a rapid cross-over from power-law 
  resistivity at high temperature to thermally activated behaviour below 
  the 120 K magnetic transition.
We do not yet have magnetic data at high pressure, which would allow us to 
  unambiguously determine $T_{MI}$, however the 
  clear signatures in $\partial\rho/\partial T$ and in 
  $\log(\rho)$ vs.\ $T$ (Fig.\ \ref{TMI}), which connect smoothly to 
  120 K at ambient pressure (Fig.\ \ref{phasediag})  
  give us confidence that we can identify $T_{MI}$.  

\begin{figure}[t]
\centering
\includegraphics[width=3.1in]{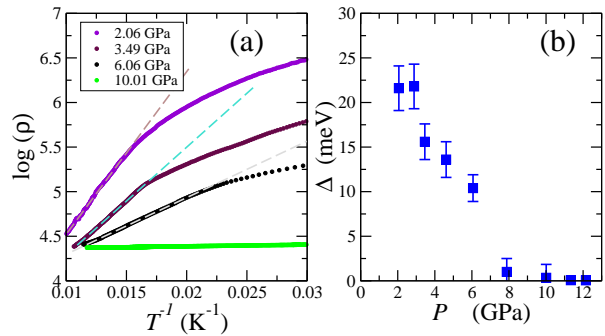}     % 
\caption{\small 
  (a) $\log(\rho(T))$ vs.\ $1/T$ at selected pressures between 100 K and 30 K.  The gap, $\Delta(P)$, 
  is extracted by fitting a single exponential to the resistivity data at 
  each pressure in the temperature range 60 K $\le T < T_{MI}$ (dashed lines). 
  These curves have been shifted vertically to allow easier comparison of 
  the slopes. 
  (b) The maximum value of the gap $\Delta(P)$, extracted as shown in figure 
    (a), is continuously suppressed by pressure. 
   } 
\label{gapvsT}
\end{figure}

\begin{figure}
\centering
\includegraphics[width=3.1in]{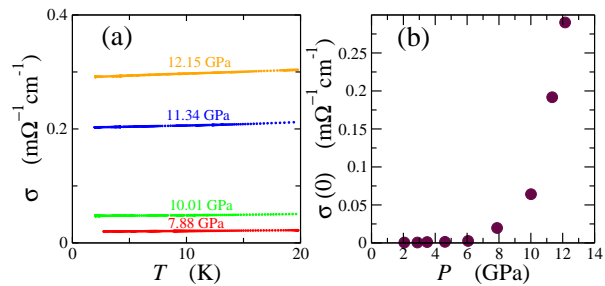}
\caption{\small 
  (a) The conductivity $\sigma(T)$ below 20 K.  The data were extrapololated to 
       0 K to get the residual conductivity values plotted in (b). 
  (b) The residual conductivity plotted as a function of pressure, 
   indicating that the pressure induced insulator to metal quantum phase transition 
   is continuous. 
}
\label{conductivity}
\end{figure}

This is further reinforced by plots of $\log(\rho)$ vs.\ $1/T$, in Fig.\ \ref{gapvsT}(a), 
  which have a straight-line dependence for a significant range of 
  temperature below $T_{MI}$ as expected 
  for a gapped system. 
If we plot the gap extracted from this straight line behaviour as a 
 function of pressure then the gap appears to vanish continuously between 
 6 and 8 GPa (Fig.\ \ref{gapvsT}(b)). 
Similarly, if we plot the $T=0$ K conductivity, $\sigma(0)$, obtained by 
 extrapolating 
 our resistivity curves to 0 K (Fig.\ \ref{conductivity}) the conductivity 
 appears to rise continuously across this pressure range. 
These behaviours are consistent with a Mott insulating ground state at 
  low pressure followed by a continuous insulator-to-metal transition between 
  6 and 8 GPa.
In this scenario, the non-divergence of the resistivity as $T\to 0$ K 
  may be due to bulk impurity states.  
Indeed, $\rho(0)$ shows a strong sample dependence, 
  consistent with impurity states in the bulk. 
However the role of disorder in the insulating phase is not clear, and 
  it should be noted that 
  the MIT at $T_{MI}$ cannot be simple disorder driven Anderson localization,  
  because disorder wipes out the insulating phase and leaves the system metallic 
  at all temperatures \cite{matsuhira07}.
Given the presence of spin 1/2 moments on the iridium sites, 
  frustration-induced localization 
  \cite{balents08} may play a role in the insulating state, 
  while the recent revelation of a commensurate AFM order in Eu$_2$Ir$_2$O$_7$ 
  below 120 K from $\mu$SR measurements raises the possibility of gapping of the 
  Fermi surface by a Slater transition \cite{zhao11}. 

% Sm$_2$Ir$_2$O$_7$ has a clear resistive signature of a thermal insulator to 
%   metal transition at 117 K \cite{matsuhira07}, and both Sm$_2$Ir$_2$O$_7$ and Eu$_2$Ir$_2$O$_7$ 
%   have magnetic phase transitions around 120 K \cite{zhao11,taira01}, 
%   so combined with the qualitative change in the temperature dependence of 
%   the resistivity (including good evidence for themally activated behaviour 
%   below around 100K -- see Fig.\ \ref{gapvsT}(b)) it is 
%   plausible that there is a metal-insulator transition in Eu$_2$Ir$_2$O$_7$ as well, 
%   even though there is no sharp feature in the resistivity.  

A difficulty with the simple Mott insulator interpretation of our phase 
  diagram is that, although 
  the gap closes continuously with pressure, 
  $T_{MI}$ does not vanish, indeed it is barely affected by pressure (Figs.\,\ref{phasediag} and 
  \ref{gapvsT}).
% This was not expected from chemical pressure experiments,
%   in which $T_{MI}$ is suppressed from 120\,K in 
%  Eu$_2$Ir$_2$O$_7$ to 36\,K in Nd$_2$Ir$_2$O$_7$ \cite{matsuhira07}, 
It is therefore interesting that recent theoretical proposals, based on band-structure 
  and many-body calculations, have variously proposed 
  strong toplogical insulator ground states \cite{pesin10,yang10,kargarian11}  
  and Weyl semi-metallic ground states 
  on the boundary of the metal-insulator transition \cite{kim11,wan11,hosur11} 
  in pyrochlore iridates.
These suggestions are consistent with our results. 
Firstly, the finite resistivity at $T \to 0$ K 
  could arise from small intrinsic Fermi pockets or surface states 
  characteristic of topological insulators. 
Moreover, Hosur et al.\ \cite{hosur11} 
  show calculated resistivity for the Weyl semi-metal that agrees qualitatively with our data.  
Indeed, the fact that $T_{MI}$ is not affected by 
  pressure, while the gap vanishes and residual conductivity rises continuously 
  across the critical pressure, could be explained if $T_{MI}$ is 
  produced by a magnetic transition of the itinerant Ir $5d$-electrons (note that, 
  according to crystal field calculations, Eu has no magnetic moment in this material) 
  that is only weakly pressure dependent, 
  while the underlying electronic structure undergoes (for example) a Lifshitz transition. 

We note that our finding of a roughly pressure independent $T_{MI}$ is in 
  contrast to a recent pressure study of Nd$_2$Ir$_2$O$_7$ \cite{sakata11} 
  which observed a monotonic suppression of $T_{MI}$ with increasing pressure. 
We do not have an explanation for this discrepancy. 
Aside from the different materials, the only obvious difference between the 
  measurements is that 
  the Nd$_2$Ir$_2$O$_7$ study used NaCl as the pressure medium, which could 
  produce anisotropic pressures which can have unpredictable effects on 
  geometrically frustrated systems.  (In this respect Daphne oil, which we 
  used, is an improvement on NaCl but still not ideal at high pressures.)

\begin{figure}
\centering
\includegraphics[width=2.7in]{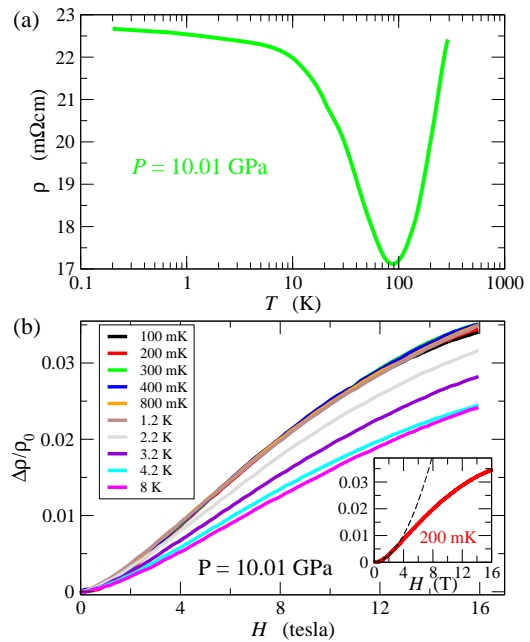}
\caption{\small 
  (a) Semi-logarithmic plot of resistivity from room temperature to 200\,mK at 
   $P=10.01$\,GPa.
  (b) Magnetoresistance as a function of magnetic field at $P = 10.01$\,GPa. 
  Data were taken in a temperature range from 100\,mK to 8\,K. 
  MR is suppressed by increasing  temperature. 
  The inset shows a quadratic fit (black dashed line, $M\propto H^2$) to the low 
  field data at $T = 200$ mK. 
  At high fields, MR tends towards saturation.}
\label{MR}
\end{figure}

After the collapse of the insulating phase above 6.06\,GPa 
  ($\Delta=0$, $\sigma(0)\neq 0$), a low-temperature NFL power law rise of resistivity survives 
  up to the highest pressure in our experiment, 
  in the negative $\partial\rho/\partial T$ metal region 
  in the bottom-right quadrant of our phase diagram. 
The nature of this phase is not clear. 
Figure \ref{MR}(a) is a semi-logarithmic plot of $\rho(T)$ at 
  $P = 10.01$\,GPa from $T = 300$\,K to 100\,mK.
The NFL resistivity behaviour fits to $\rho(T)=\rho(0)-AT^x$ with 
  $x=1.0\pm0.1$ below 20\,K for $P > 6$\,GPa.
While the resistivity upturn of the metallic 
  phase below $T_{min}$ looks Kondo-like, similar to what is observed in the frustrated 
  Kondo lattice of Pr$_2$Ir$_2$O$_7$ \cite{nakatsuji06}, 
  crystal-field analyses that predict no local $f$ moments  
  in Eu$_2$Ir$_2$O$_7$ \cite{machida05}  would seem to rule out 
  a Kondo effect.
Similarly, recent theoretical work \cite{udagawa12} that ascribes the 
 resistive upturn in Pr$_2$Ir$_2$O$_7$ and in Nd$_2$Ir$_2$O$_7$ at high 
 pressure \cite{sakata11} to frustrated spin-ice-like correlations among 
 the local $f$-electrons, would also appear to be ruled out by the absence of 
 $f$-moments on the Eu sites.  

As well as being a common feature of the metallic pyrochlore iridates, 
  a rising resistivity as $T\rightarrow 0$ K 
  has also been observed in 
  metallic pyrochlore molybdates  at high pressures. 
Hydrostatic pressure destroys the FM order in the metallic ground states of 
  (Nd and Sm)$_2$Mo$_2$O$_7$ by tuning the relative strength of double and 
  super-exchange interactions amongst Mo $d$ electrons. 
The resulting order-disorder transition coincides with a change in the 
  resistivity behaviour, from conventional metallic to a ``diffusive" NFL state 
  \cite{iguchi09} characterized by a large, relatively weakly $T$ dependent 
  resistivity, not unlike the resistivity of our $\partial\rho/\partial T > 0$ 
  metal, and also having an upturn in the resistivity as $T\rightarrow 0$ K. 
In this context, it seems significant that 
  the tetragonal antiferromagnet Ba$_2$IrO$_4$ \cite{okabe11} 
  has recently been shown to have a continuous pressure-induced insulator-to-metal transition  
  that is related to suppression of magnetism, but 
  in this case the low temperature resistivity at high pressure is metallic. 
Thus, the upturn in $\rho(T)$ as $T\rightarrow 0$ K is not a generic feature of 
  iridates near the boundary of a Mott insulating ground state, rather  
  the pyrochlore lattice structure seems to play a decisive role. 
%%%%%%%%%%%%%%%%%%%%%%%%%%%%%%%%%%%%%%%%%%%%%%%%%%%%%%

To further investigate the negative $\partial\rho/\partial T$ metallic phase, 
  we measured magnetoresistance 
  (MR) at P = 10.01\,GPa by sweeping magnetic field from 0 to 16 tesla at ten 
  different temperatures from 100\,mK to 8\,K (Fig.\,\ref{MR}). 
The MR signal is positive, which rules out weak localization as the cause of 
  the low temperature upturn in $\rho(T)$, as had been suggested in the 
  molybdates \cite{iguchi09}.
It probably also rules out other mechanisms involving scattering from spins \cite{udagawa12}, 
  and indeed it may constrain the possible magnetic order on the Ir sublattice. 
According to reference \cite{wan11}, if the Ir moments are ferromagnetically  
  aligned the ground state will be metallic; 
  FM metals usually have a negative MR however\cite{pippard89}, in contrast to our
  observations of a positive signal. 
So our results suggest that the 
  ground state is either antiferromagnetic or has the so-called ``all-in/all-out" 
  configuration of spins on each Ir tetrahedron.  
According to references \cite{wan11,kim11}, these 
  magnetic configurations could have a Weyl semi-metallic phase separating 
  the Mott insulating and the metallic ground states.

The weak positive MR signal grows quadratically at low fields ($M\propto H^2$), 
  tends towards saturation at high fields, and becomes smaller with 
  increasing temperature. 
Such behaviour is generic to (non-ferromagnetic) metals with closed orbits 
  on the Fermi surface \cite{pippard89}. 
The weakness of the  MR signal (Fig.\,\ref{MR}) may be due to strong 
  scattering, otherwise a state with small pockets should have large MR due to the small 
  Fermi volume and the low density of carriers. 
The origin of such strong scattering is not clear, although it is consistent with 
  the fairly large resistivity of these samples, even in the metallic regime; 
  moreover the metallic state must be unconventional due to the non-metallic slope of 
  the resistivity $\partial\rho/\partial T < 0$, and the non-Fermi-liquid 
  power-law $T$ dependence.

Further measurements are urgently required in order to elucidate the nature of 
  the negative $\partial\rho/\partial T$ metallic phase. 
For example, reference \cite{wan11} predicts that
  the all in/out spin configuration
  on the Ir tetrahedra has a 1.3 meV gap to the collinear ferromagnetic
  state. 
With this gap, our maximum field of 16 T could be sufficient to cause a 
  reorientation of the spins to ferromagnetic, which would cause a huge 
  jump in magnetoresistance, which we do not observe.
However, depending on the accuracy of that calculation,
  a higher magnetic field may be required, so 
  higher field measurements would be informative.

\section{Conclusion}

Using resistivity, we have mapped the temperature-pressure phase diagram of 
  Eu$_2$Ir$_2$O$_7$ between 2 and 12 GPa. 
The metal-insulator boundary is near 6 GPa.  
At high temperature ($T>100$ K), the resistivity falls by a factor of more than 
  60 between 2 and 12 GPa, and the behaviour crosses over near 6 GPa from an incoherent metal 
  with a very high resistivity and a negative $\partial \rho/\partial T$, 
  to a more conventional metal having a positive $\partial \rho/\partial T$.
At intermediate pressures the cross-over is observed as a function of temperature: 
  the sign of $\partial \rho/\partial T$ changes from positive to negative at 
  a temperature $T^*$ that was not observed using chemical pressure.

The low temperature behaviour evolves from a low pressure ``insulating" state, having a 
  temperature dependent gap but a resistivity that does not diverge as $T\rightarrow 0$ K, 
  to an anomalous metallic state again having a negative $\partial \rho/\partial T$. 
Conventional explanations for this rising resistivity as $T \rightarrow 0$ K, such 
  as the Kondo effect or weak localization, are ruled out.  
The transition between these ground states is continuous, 
  with the critical pressure near 6 GPa, however there is 
  no temperature scale apparent in the resistivity that vanishes at the critical pressure, 
  rather the high-to-low temperature cross-over persists across the entire 
  phase diagram.  
An obvious scenario is that the low-temperature phases occur below a 
  magnetic phase transition that is essentially unaffected by pressure. 

The high absolute value and the low temperature rise of the resistivity in the 
  negative $\partial \rho/\partial T$ metallic phase of Eu$_2$Ir$_2$O$_7$ might be 
  manifestations of a topological semimetallic phase. 
The anomalous nature of the ground states on both sides of the QCP suggests 
  a topological character for the quantum phase transition and may explain its unusual form. 
These results should encourage further experiments to test for the existence of 
  topological states near the metal-insulator boundary of the pyrochlore iridates. 

\begin{acknowledgments}
The authors acknowledge helpful discussions with Y. B. Kim, Y. Machida, 
  W. Witczak-Krempa, A. Ramirez and V. Dobrosavljevic. FFT is grateful to Mark Aoshima for his 
  invaluable machining skills. 
This research was supported by the Natural Science and Engineering Research Council 
  of Canada, the Canadian Institute for Advanced Research, a Grant-in-Aid (No. 21684019) 
  from JSPS, and a Grant-in-Aid for Scientific Research on Priority Areas 
  (19052003) from MEXT, Japan.
\end{acknowledgments}

\bibliography{fazel}% Produces the bibliography via BibTeX.

\end{document}